\documentclass[conference]{IEEEtran}
\IEEEoverridecommandlockouts
\usepackage{cite}
\usepackage{amsmath,amssymb,amsfonts}
\usepackage{graphicx}
\usepackage{textcomp}
\usepackage{xcolor}

\usepackage{multirow}
\usepackage{algorithmicx}
\usepackage{algpseudocode} 
\usepackage{booktabs}
\usepackage[most]{tcolorbox}
\usepackage[table]{xcolor}
\usepackage{soul}
\usepackage{xurl} 
\usepackage{twemojis}

\def\BibTeX{{\rm B\kern-.05em{\sc i\kern-.025em b}\kern-.08em
    T\kern-.1667em\lower.7ex\hbox{E}\kern-.125emX}}
\begin{document}

\title{RotDroid: Cross-Orientation State Equivalence Testing for Detecting GUI Rotation Bugs in Android Apps}

\author{
\IEEEauthorblockN{Mengdi Qin}
\IEEEauthorblockA{
\textit{SKLCCSE} \\
\textit{Beihang University} \\
Beijing 100191, China \\
qinmengdi@buaa.edu.cn
}
\and
\IEEEauthorblockN{Bo Jiang\textsuperscript{*}}
\IEEEauthorblockA{
\textit{SKLCCSE} \\
\textit{Beihang University} \\
Beijing 100191, China \\
jiangbo@buaa.edu.cn
}
\thanks{\textsuperscript{*}Bo Jiang is the corresponding author.}
}

\maketitle

\begin{abstract}
Screen rotation is a fundamental interaction in Android applications, but it often introduces non-crashing functional failures (NCFs), such as layout inconsistencies and state loss, which are difficult to detect automatically. A key challenge is the lack of effective test oracles for checking cross-orientation state equivalence between portrait and landscape views. We propose RotDroid, a testing framework for detecting GUI rotation bugs via cross-orientation state equivalence. RotDroid generates and mutates State-Preserving action Sequences (SPS) to construct semantically equivalent GUI states across orientations. To support reliable oracle checking, we build RotBench, a dataset of paired portrait–landscape GUI states, and develop RotVL, a vision-language model fine-tuned for equivalence checking. Experiments on both synthetic and real-world datasets show that RotVL outperforms state-of-the-art models, and RotDroid detects more rotation-induced failures than existing techniques under equal budgets. In large-scale studies on open- and closed-source apps, RotDroid reports 94 previously unknown bugs, with 47 confirmed or fixed by developers, demonstrating its practical effectiveness.

\end{abstract}


\section{Introduction}

Screen orientation change is a common form of runtime configuration change in mobile applications. With recent Android policies mandating orientation flexibility\cite{android16_orientation_resizability_2025}, applications are required to support seamless layout adaptation across portrait and landscape modes. However, correct adaptation remains challenging due to the Android lifecycle mechanism, where orientation changes trigger a destroy-and-recreate process that requires developers to manually preserve transient states\cite{amalfitano2018does}. As a result, orientation changes frequently introduce non-crashing functional (NCF) defects\cite{amalfitano2018does,ju2024study}, such as component occlusion \cite{liu2022nighthawk} or data loss \cite{riganelli2020data-dld,guo2022ifixdataloss}. Unlike crashes, these defects manifest as silent functional impairments, making them difficult to detect with existing testing techniques.

Existing automated testing approaches have limited effectiveness in detecting such GUI rotation bugs. GUI exploration tools primarily target crashes and overlook visual logic errors, while vision-based approaches\cite{liu2022nighthawk} focus on anomalies within a single static view and cannot capture cross-orientation inconsistencies. Techniques based on structural similarity \cite{fazzini2017automated-diffdroid} are inapplicable due to differing screen layouts across orientations. Methods that validate state restoration after consecutive  orientation changes (DOC) (i.e., $Portrait \rightarrow Landscape \rightarrow Portrait$)  \cite{amalfitano2018does} miss defects in intermediate states, and data-loss detectors \cite{riganelli2020data-dld,guo2022ifixdataloss} fail to identify layout-related issues. Although recent multimodal approaches \cite{liu2025seeing-visiondroid} leverage large vision-language models, they lack the domain-specific knowledge required to distinguish valid layout adaptation from rotation-induced defects.

In summary, detecting GUI rotation bugs faces two key challenges. First, designing an effective oracle for cross-orientation state equivalence is difficult, as it requires accurately determining whether portrait and landscape views represent the same semantic state. Second, rotation bugs are difficult to expose through naive orientation switching alone, as they often manifest only when orientation changes are interleaved with user interactions.

To address these challenges, we propose a systematic framework for cross-orientation GUI testing. We first conduct an empirical study to characterize GUI rotation bugs, which informs the construction of RotBench, a dataset of both equivalent and non-equivalent portrait–landscape GUI pairs. Based on this dataset, we develop RotVL, a vision-language model for cross-orientation state equivalence checking. Building on this oracle, we design RotDroid, an automated testing framework that generates and mutates state-preserving action sequences (SPS) to construct semantically equivalent GUI states across orientations. RotDroid then leverages RotVL to analyze these paired states and detect rotation-induced inconsistencies.

Our results show that RotVL-8B outperforms state-of-the-art vision-language models, including Qwen3-VL-235B-A22B-Instruct and GPT-5.2, on cross-orientation bug detection tasks. In large-scale evaluations on 300 open-source and 103 commercial Android applications, RotDroid reports 72 previously unknown bugs in open-source apps (41 confirmed and 29 fixed) and 22 previously unknown bugs in commercial apps (6 confirmed), demonstrating its effectiveness in detecting real-world GUI rotation issues.

In summary, the contributions of this work are as follows:
\begin{table*}[!h]
\centering
\caption{Distribution of rotation-induced mobile GUI issues.}
\label{tab:defects}
\footnotesize
\renewcommand{\arraystretch}{1} 
\resizebox{\textwidth}{!}{
    \begin{tabular}{llcl} 
    \toprule
    \textbf{Category} & \textbf{Bug Type} & \textbf{Percentage} & \textbf{Description} \\
    \midrule
    \multirow{2}{*}{\shortstack[l]{Rotation Execution Failures}} 
     & Unintended Rotation & 18\% & 
     The screen or UI elements rotate unnecessarily when should remain fixed.\\
     & No Rotation & 16\% & The screen or UI elements fail to rotate when the device orientation changes.\\
    \midrule
    \multirow{3}{*}{\shortstack[l]{Layout Failures}} 
     & Content Clipping & 14\% & Parts of the content or components are truncated or extend beyond the visible screen. \\
     & Component Absence & 12\% & Essential UI elements are not displayed or rendered correctly. \\
     & Component Overlap & 9\% & UI components or contents overlap or obscure each other.\\
    \midrule
    \multirow{3}{*}{\shortstack[l]{State Restoration Failures}} 
     & Navigation State Loss & 12\% & The scroll position, pagination, or other navigational state return to default position. \\
     & Interrupted Activities & 8\% & Ongoing activities, such as video playback or animations, stop or restart unexpectedly. \\
     & User Input Loss & 3\% & Form or input field data is lost. \\
    \midrule
    Other & Other & 8\% & Hardware-specific display issues, performance glitches, or other rare symptoms. \\
    \bottomrule
    \end{tabular}
}
\end{table*}

\begin{itemize}
\item We conduct an empirical study to systematically characterize GUI rotation bugs and derive a taxonomy that informs the design of our approach.

\item We introduce RotBench, a dataset of paired portrait–landscape GUI states with both equivalent and non-equivalent cases, and develop RotVL, a vision-language model to support cross-orientation state equivalence checking.

\item We propose RotDroid, an automated testing framework that generates and mutates state-preserving action sequences (SPS) to construct semantically equivalent GUI states across orientations and detect rotation-induced inconsistencies.

\item We evaluate RotDroid on large-scale open-source and commercial Android applications, where it identifies numerous previously unknown GUI rotation bugs, many of which have been confirmed or fixed by developers.
\end{itemize}

\section{Empirical Study}
\label{sec:empirical study}

Unlike the prior study on DOC GUI failures~\cite{amalfitano2018does}, which focuses on failures exposed by two consecutive orientation changes and same-orientation GUI comparison, our study mines real-world issue reports and derives a symptom-oriented taxonomy of rotation-induced GUI bugs. This taxonomy guides RotBench construction and motivates RotDroid's portrait--landscape consistency checking.

We mined public GitHub issue reports using orientation-related keywords such as ``app rotation'' and ``screen orientation''. To improve relevance, we restricted the scope to closed issues reported within the past five years, obtaining 6,031 unique reports.

Two authors manually inspected each report to determine whether it described a mobile GUI issue and whether the issue was explicitly triggered by screen rotation. Disagreements were resolved through discussion until consensus was reached. This process reduced the initial 6,031 reports to 898 confirmed rotation-induced GUI issues.

Before large-scale labeling, two authors independently analyzed a pilot set of 100 reports and iteratively derived an initial taxonomy based on user-observable symptoms. We organize rotation-induced failures into four high-level families: Rotation Execution Failures, Layout Failures, State Restoration Failures, and Other. Since one report may mention multiple symptoms, we assign each report a single primary label according to the dominant user-perceivable failure. The inter-rater agreement before discussion was 94.2\%, with Cohen's kappa of 0.89, indicating strong agreement.

Table~\ref{tab:defects} summarizes the final distribution. Layout Failures, Rotation Execution Failures, and State Restoration Failures account for 35\%, 34\%, and 23\% of the reports, respectively, while the remaining 8\% fall into Other, mainly hardware-specific rendering issues, performance glitches, and rare symptoms that cannot be reliably synthesized. RotBench therefore focuses on the five dominant defect types that can be systematically generated and validated.

\section{The RotVL Model}
\label{sec:RotVL}

Since existing Vision-Language Models (VLMs) are ineffective for checking state equivalence within paired portrait--landscape GUI states, 
we propose RotVL, a \textbf{Rot}ation-Aware \textbf{V}ision–\textbf{L}anguage Model fine-tuned for detecting cross-orientation GUI inconsistencies. 
Our methodology is built upon two core pillars: First, we construct the Pairwise Rotation-related GUI Defect Dataset (RotBench). It combines non-defective pairs captured via an automated equivalent GUI state pairs collection pipeline with diverse defective pairs generated through data augmentation. Second, we adopt a LoRA-based Supervised Fine-tuning strategy based on Qwen3-VL \cite{bai2025qwen3vltechnicalreport}. We employ a two-task unified training approach that enables the model to determine the state equivalence between paired portrait--landscape snapshots.

\subsection{RotBench: Pairwise Rotation-related GUI Defect Dataset}
\label{subsec:RotBench}

\subsubsection{Automated Collection of Non-defective Pairs} 
\label{sec:rightpairs}
\begin{figure}[!htbp]
\centering
\includegraphics[width=0.9\linewidth]{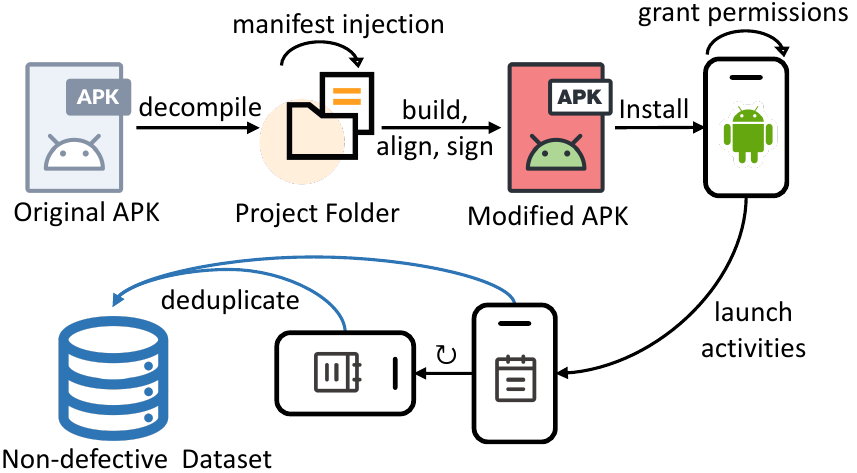}
\caption{Pipeline of the Non-defective Pairs Collection.}
\label{fig:rightdata}
\end{figure}

We collected a set of 851 candidate applications from F-Droid~\cite{fdroid}. Using these candidates as input, we developed an automated pipeline to construct the non-defective pairs dataset, as illustrated in Figure \ref{fig:rightdata}. 

\begin{figure*}[!htbp]
\centering
\includegraphics[width=0.9\textwidth]{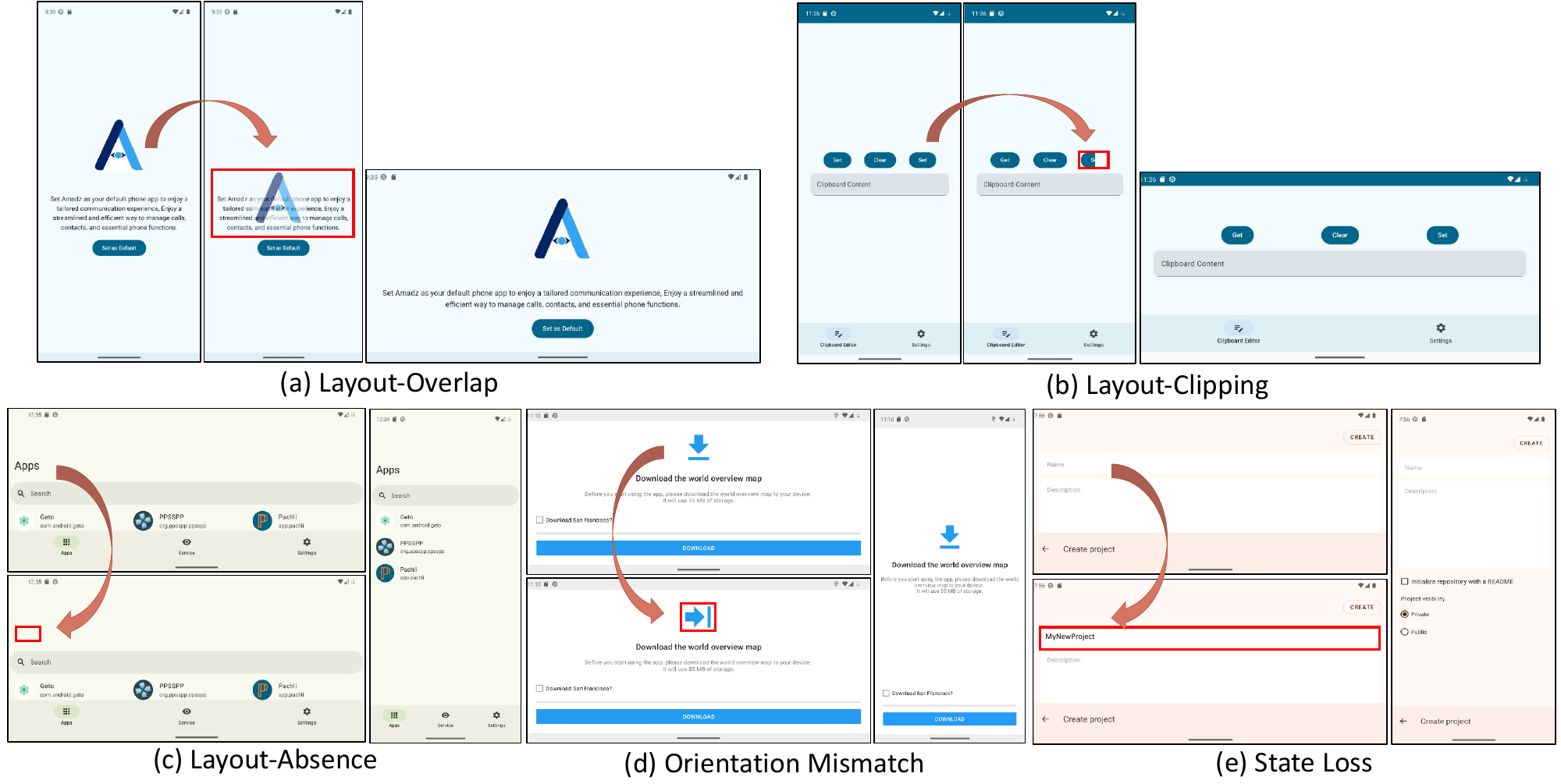}
\caption{Examples of Automated Generation of Defective Pairs.}
\label{fig:buggydata}
\end{figure*}
First, we decompile the applications to modify the \texttt{AndroidManifest.xml} file. 
This process involves exporting all activities and injecting runtime permissions to prevent system dialog interruptions. We subsequently rebuild, align, and sign the modified application for deployment.
Second, we install the modified APKs and systematically launch each target activity and capture each GUI state (view hierarchy and snapshot) in both portrait and landscape orientations. 
Third, we apply a strict filtration process to ensure data quality. 
We remove invalid pairs caused by rotation failures by comparing resolution dimensions. 
Additionally, we eliminate visual redundancies using Difference Hash algorithm. 
A final manual verification removes any pre-existing GUI defect artifacts.
Consequently, this process yielded a refined dataset comprising 1,818 valid pairs derived from 623 applications.

\subsubsection{Automated Generation of Defective Pairs}
Having collected the non-defective pairs, we perform data augmentation to synthesize datasets of defective pairs. 
The pipeline accepts paired portrait and landscape GUI screenshots alongside their view hierarchies (in XML) as input. 
The output consists of screenshot pairs where one orientation contains an injected defect. Each pair is associated with a label that encapsulates the defect category, the affected orientation, and the precise bounding box coordinates of the anomaly.
The generation process consists of three sequential stages.

First, the framework parses the XML metadata (view hierarchy) to extract candidate UI widgets. 
To ensure consistency, we verify that the selected widgets are visible and persistent across both orientation states through coordinate mapping and content matching. 
Second, we apply distinct mutation operators to the valid candidate regions to reproduce the five representative defect types, as exemplified in Figure~\ref{fig:buggydata}. 
Specifically, we simulate \textit{Component Absence} and \textit{Content Clipping} by masking the entire or partial widget area with boundary color. \textit{Component Overlap} is synthesized by relocating a source widget onto a target region using transparency blending. \textit{Orientation Mismatch} is implemented via affine transformations to rotate local regions by $90^{\circ}$ or $180^{\circ}$. Finally, \textit{State Loss} is reproduced by erasing existing input text and substituting it with synthetic content generated by a Multimodal Large Language Model (MLLM).
Finally, we employ a post-generation validation process to guarantee data quality.
We quantify the visual discrepancy between the original and mutated states using Structural Similarity Index (SSIM) or Optical Character Recognition (OCR) analysis. 
The system automatically discards any samples that fail to exhibit statistically significant perceptible differences.

Upon completion of the data augmentation, we successfully constructed a large-scale dataset, with a total of 11,233 paired samples. The distribution comprises 2,950 Component Absence, 2,852 Component Overlap, 2,783 Content Clipping, 2,024 Orientation Mismatch, and 624 State Loss instances.

\begin{table*}[!htbp]
  \centering
  \caption{Statistics of the Constructed RotBench Dataset for Bug Detection, Classification, and Localization.}
  \label{tab:rq1}
  \begin{tabular}{lcccccccc}
    \toprule
    \multirow{2}{*}{\textbf{Task}} & \multicolumn{2}{c}{\textbf{Training Set}} & \multicolumn{2}{c}{\textbf{Validation Set}} & \multicolumn{2}{c}{\textbf{Test Set}} & \multicolumn{2}{c}{\textbf{Total}} \\
    \cmidrule(lr){2-3} \cmidrule(lr){4-5} \cmidrule(lr){6-7} \cmidrule(lr){8-9}
     & \# Samples & \# Apps & \# Samples & \# Apps & \# Samples & \# Apps & \# Samples & \# Apps \\
    \midrule
    Bug Detection   & 2,960 & 502 & 320 & 59 & 340 & 62 & 3,620 & 623 \\
    Bug Classification and Localization & 2,300 & 425 & 320 & 51 & 340 & 54 & 2,960 & 530 \\
    \bottomrule
  \end{tabular}
\end{table*}

\subsubsection{Data Balancing and Partitioning Strategy}
To ensure the generalization capability of the model over unseen applications, the partitioning mechanism operates strictly at the project level rather than the sample level. We divide the dataset into training, validation, and testing subsets using an 8:1:1 ratio based on unique project identifiers. This strategy guarantees that visual patterns or defect characteristics specific to a single project do not appear simultaneously in the training and evaluation sets. 

Following the project-wise split, inherent disparities in project sizes and bug frequencies introduce class imbalance.
Specifically, the ``State Loss'' category, which relies on specific \texttt{EditText}-based mutations, inherently constitutes the minority class compared to layout-based defects. To mitigate this, we employ a fine-grained Stratified Under-sampling Strategy. We define the set of defect types $\mathcal{T}$ and screen orientations $\mathcal{O}$ as:
\begin{equation}
\mathcal{T} = \{ \text{State Loss, Overlap, Clip, Absence, Orientation} \}
\end{equation}
\begin{equation}
\mathcal{O} = \{ \text{Portrait, Landscape} \}
\end{equation}

The balancing process operates in two stages based on the downstream task:
\begin{enumerate}
\item Cross-Attribute Normalization: First, we categorize all defective samples in the training pool into subgroups $\mathcal{S}_{(t,o)}$ based on the Cartesian product of defect type $t \in \mathcal{T}$ and orientation $o \in \mathcal{O}$. To ensure robustness against both defect variety and device rotation, we calculate a balancing threshold $\kappa$ determined by the minority subgroup count:
\begin{equation}
\kappa = \min_{t \in \mathcal{T}, o \in \mathcal{O}} |\mathcal{S}_{(t,o)}|
\end{equation}

All majority subgroups are randomly undersampled to match size $\kappa$.

\item Task-Specific Dataset Construction: Based on the normalized subgroups, we construct two distinct datasets:
\begin{itemize}
    \item \textit{For Bug Detection:} The goal is to distinguish between defective and non-defective screens. 
    Let ${N}_{clean}$ represent the total number of non-defective samples. Let $M$ denote the total number of subgroups where $M = \mathcal{|T| \times |O|}$. 
    We randomly sample an equivalent number of non-defective and defective screens to form the negative set $\mathcal{D}_{neg}$ and the positive set $\mathcal{D}_{pos}$, satisfying:
    \begin{equation}
    |\mathcal{D}_{pos}| = |\mathcal{D}_{neg}| = M \times  \min \left(\kappa, \left\lfloor \frac{{N}_{clean}}{M} \right\rfloor \right)
    \end{equation}
    This 1:1 ratio prevents the model from biasing towards the majority non-defective class.
    
    \item \textit{For Bug Classification and Localization:} The objective is to categorize the specific defect type and localize the visual anomaly. The dataset is constructed strictly from defective samples. We maintain the equilibrium established in the normalization phase, ensuring that all five defect classes are represented equally:
    \begin{equation}
    |\mathcal{C}_{t}| = 2\kappa, \quad \forall t \in \mathcal{T}
    \end{equation}
\end{itemize}
\end{enumerate}

Consequently, the final RotBench dataset, constructed via these rigorous generation and balancing procedures, is summarized in Table~\ref{tab:rq1}.

\subsection{Fine-tuning Strategy}
\subsubsection{Task-Aware Instruction Design}

To effectively facilitate the LoRA-based fine-tuning process, we designed a structured, two-stage instruction template that guides the model for effective bug detection and diagnosis. This coarse-to-fine strategy mimics the cognitive process of human testers and mitigates the risk of hallucination in non-defective samples. As illustrated in Figure \ref{fig:prompt}, the instruction design consists of the following key components:

\begin{figure}[!htbp]
\centering
\includegraphics[width=0.95\columnwidth]{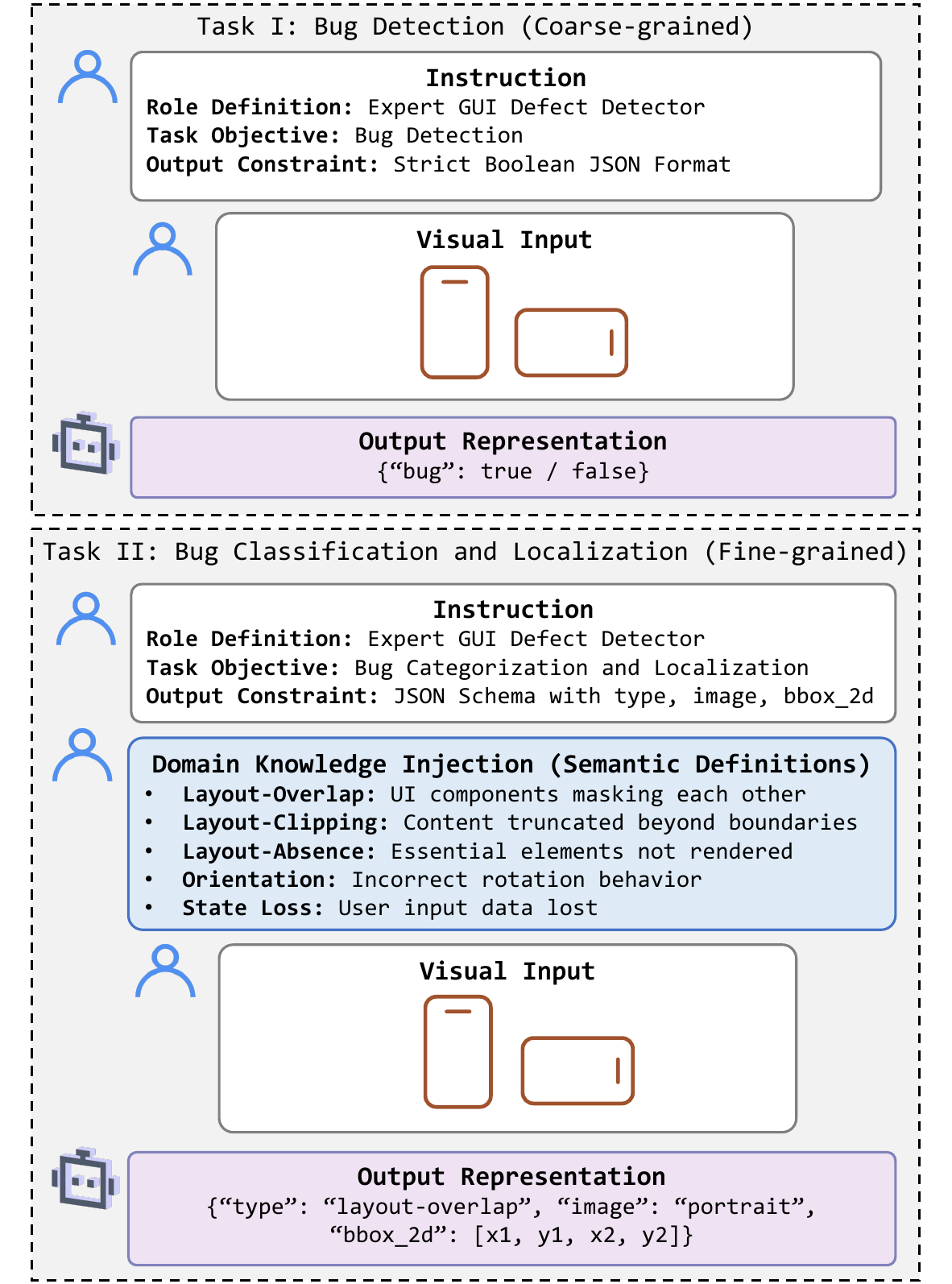}
\caption{Prompt for Bug Detection, Classification, and Localization.}
\label{fig:prompt}
\end{figure}

\begin{figure*}[t]
\centering
\includegraphics[width=1\textwidth]{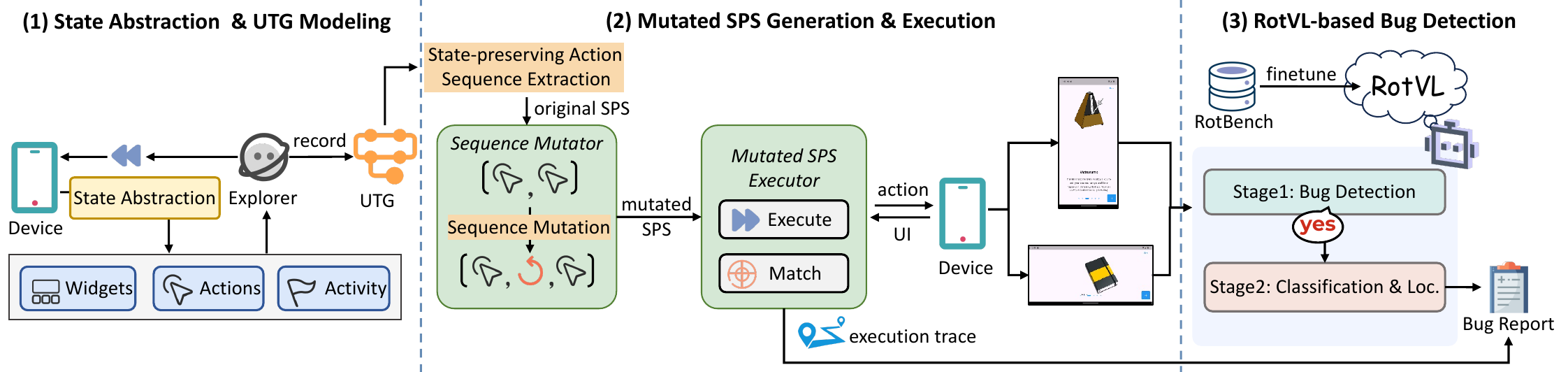}
\caption{Overview of the RotDroid.}
\label{fig:overview}
\end{figure*}

\begin{itemize}
    \item \textbf{Role Definition and Task Decomposition:} We explicitly define the model's role as an ``Expert GUI defect detector'' to prime its latent space for domain-specific tasks. The complex objective is decomposed into two sub-tasks: (1) \textit{Bug Detection}, a binary classification task to determine the presence of defects, and (2) \textit{Bug Classification and Localization}, a dense prediction task learned exclusively from positive samples.
    
    \item \textbf{Context-Rich Semantic Definitions:} In the bug classification and localization stage, rather than merely providing class labels, we embed detailed semantic definitions for all five defect categories (e.g., explaining \texttt{layout-overlap} as ``UI components masking each other'') directly into the prompt. This injects domain knowledge and helps the model distinguish between visually similar defects (e.g., \texttt{layout-clipping} vs. \texttt{layout-absence}).
    
    \item \textbf{Structured Output Constraints:} To ensure the model's output is parsing-friendly for downstream evaluation, we enforce a strict JSON schema. 
     For detection, the output is restricted to a boolean indicator (i.e., \texttt{\{"bug": true/false\}}). For classification and localization, the model is instructed to predict the defect's category (\texttt{type}), the affected screen orientation (\texttt{image}), and the spatial location (\texttt{bbox\_2d}).
\end{itemize}

During training, we format the ground truth data into this conversational format, masking the user instructions and calculating loss only on the model's JSON responses.

\subsubsection{Training Strategy and Optimization} \label{subsubsec:training}
We employ the Qwen3-VL as our backbone model. 
To achieve efficient adaptation while preserving the pre-trained generalization capabilities of the model, we apply Low-Rank Adaptation (LoRA) to the linear layers. 
We perform unified fine-tuning on the RotBench dataset to simultaneously optimize the model for bug detection, classification and localization tasks. 
Specifically, we formulate both tasks as next-token prediction problems using a standardized instruction-following format. 
We use ZeRO-3 Offload to reduce GPU memory consumption by offloading training states to host memory, while using AdamW as the optimizer and a cosine annealing schedule to adjust the learning rate.

\section{RotDroid}
\label{sec:RotDroid}
In this section, we present RotDroid, an automated framework designed to systematically detect and diagnose GUI rotation bugs in Android apps. As illustrated in Figure \ref{fig:overview}, the framework comprises three modules: (1) UI Transition Graph Modeling, which systematically explores the application to construct a structure-aware GUI state transition model; (2) Mutated SPS Generation and Execution, which synthesizes state-equivalent portrait--landscape screen pairs by extracting and mutating SPS from the constructed graph; and (3) RotVL-based Bug Detection, which leverages RotVL to detect and diagnose GUI rotation bugs. The following subsections detail the design and implementation of each component.

\subsection{State Abstraction and UI Transition Graph Modeling}

\subsubsection{Robust State Abstraction}
First, regarding action identification, we extract interactive widgets from the view hierarchy, excluding nodes with \texttt{enabled}=\texttt{false} or \texttt{visible-to-user}=\texttt{false}. For composite layouts, \textit{recursive semantic aggregation} fills each missing \texttt{text}, \texttt{hint}, or \texttt{content-desc} attribute with the first non-empty descendant value, without concatenating text or aggregating semantics across siblings. We also calculate a robust locator (e.g., XPath) for each element, so every action has semantic and structural descriptors.
Second, regarding state definition, we map the runtime environment to a unique signature by combining the current Activity context with the structural layout of visible components. We explicitly filter out volatile system-level noise (e.g., time, network status) to prevent state explosion, ensuring that functionally identical screens remain deterministically consistent.

\subsubsection{UI Transition Graph Construction}
We follow previous work \cite{li2017droidbot,sustoat} to build a UI Transition Graph (UTG) to model the application's execution behavior. Each node represents all the information for a GUI state, including the states of all its widgets and their hierarchical relationship. Each edge represents the action that causes a transition from the source node to the target node.

\subsection{Mutated SPS Generation and Execution}

To systematically detect inconsistencies across different screen orientations, we propose a method based on \textbf{S}tate-\textbf{P}reserving action \textbf{S}equences (SPS). We define a state-preserving action sequence as an action sequence $\mathcal{A}= \langle a_1, a_2, \dots, a_n \rangle$, where an initial portrait state, $S^p_1$, undergoes a sequence of actions $\mathcal{A}$ and ultimately returns to a portrait state $S^p_n$ that is semantically equivalent to $S^p_1$, that is, operationally equivalent in observable GUI functionality and user-visible semantics, rather than formally equivalent in all hidden application state. We denote the portrait target state before mutation $S^p_1$ as $S_{target}^{p1}$.
To construct state-equivalent portrait--landscape pairs, we insert rotation actions into the SPS. By injecting a rotation action $a_{rot}$ into $\mathcal{A}$, we generate a mutated sequence $\mathcal{A}' = \langle a_1, \dots, a_k, a_{rot}, a_{k+1}, \dots, a_n \rangle$. Executing $\mathcal{A}'$ drives the application to a corresponding landscape target state $S^l_n$, denoted as $S_{target}^{l}$. We then insert another rotation action after $S_{target}^{l}$ to rotate the app back to portrait mode, obtaining a second portrait target state $S_{target}^{p2}$. 
Consequently, we form two state-equivalent portrait--landscape pairs for differential testing: $(S_{target}^{p1}, S_{target}^{l})$ and $(S_{target}^{p2}, S_{target}^{l})$.
This additional rotation does not reduce the oracle to DOC-style same-orientation checking. DOC-based methods~\cite{amalfitano2018does} perform consecutive portrait--landscape--portrait rotations without interleaved user actions and compare only the initial and final portrait states. In contrast, RotDroid inserts rotation into an SPS and explicitly checks the intermediate landscape state $S_{target}^{l}$ against both portrait target states.

\subsubsection{State-preserving Action Sequence Extraction}
\label{sec:srs_extraction}
The acquisition of valid SPS is based on the UTG. In this context, an SPS manifests as a closed-loop execution trajectory, where the system transitions from a specific state $S_{start}$ and returns to it via a sequence of interactions.

The extraction process is triggered whenever a new transition edge $e_{new}$ is added to the UTG. The system employs a depth-first search (DFS) algorithm to identify all simple cycles containing $e_{new}$. To ensure these abstract cycles are executable from the application's entry point, we perform a reachability analysis. For each identified cycle $C$ starting at state $S_{start}$, we compute the shortest path $P_{init}$ from the application's entry node $S_{root}$ to $S_{start}$. The final executable sequence is constructed by concatenating the initialization path with the cycle: $\mathcal{S} = P_{init} \oplus C$. This integration ensures that the testing agent can deterministically navigate to the target context before verifying the state recurrence.

\subsubsection{State-preserving Action Sequence Mutation}
\begin{figure*}[!htbp]
\centering
\includegraphics[width=0.95\textwidth]{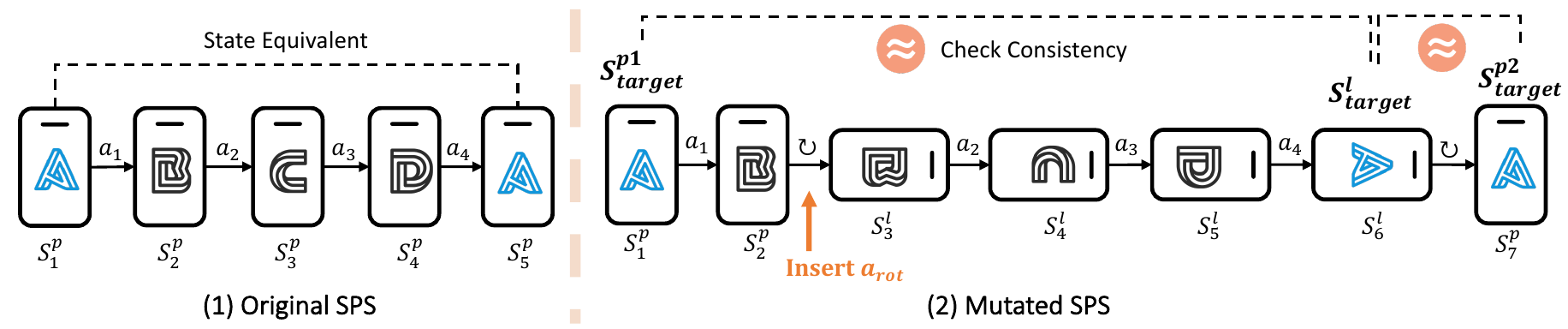}
\caption{State-equivalent Portrait--Landscape Pairs Construction via State-preserving Action Sequence Mutation.}
\label{fig:srsmutation}
\end{figure*}

To construct the state-equivalent portrait--landscape pairs required for cross-orientation consistency testing, we apply a mutation operator to the acquired SPS. The core principle is to introduce a screen orientation change, a rotation action $a_{rot}$, into the established recovery sequence, thereby forcing the application to reconstruct the UI layout under a different configuration, as illustrated in Figure \ref{fig:srsmutation}.

Given a base sequence $\mathcal{A} = \langle a_1, \dots, a_n \rangle$, the mutation process generates a set of variant sequences $\mathbf{\Omega}$.
To construct a diverse set of state-equivalent portrait--landscape pairs, we employ random sampling without replacement to select a subset of unique insertion indices from the range $[0, n)$. For each sampled index $k$, we inject a rotation action $a_{rot}$ to generate a distinct mutated sequence.
This mutation strategy transforms the original state recovery task into cross-orientation consistency checks, where the landscape target state $S_{target}^{l}$ is expected to be semantically equivalent to both the original portrait target state $S_{target}^{p1}$ and the rotated-back portrait target state $S_{target}^{p2}$.

\subsubsection{Mutated State-preserving Action Sequence Execution}
\label{sec:srs_execution} 
Executing the mutated SPS presents a significant challenge due to the dynamic nature of Android GUIs, where runtime widget attributes (e.g., list contents or memory addresses) may differ from the static snapshot stored in the UTG.
To address this, we implement a robust \textit{Adaptive Action Execution} mechanism with a hierarchical spatial matching strategy.

Before executing a planned action $a_{plan}$ targeting a recorded widget $w_{rec}$, the system captures the current runtime hierarchy and attempts to bind $w_{rec}$ to a real-time element $w_{cur}$. The matching logic follows a cascading priority protocol:

\begin{enumerate}
    \item \textbf{Unique Identifier Match:} Priority is given to elements with identical invariant \texttt{resource-id}.
    \item \textbf{Semantic Attribute Match:} If IDs are absent or ambiguous, the system matches by class name and aggregated semantic attributes.
    \item \textbf{Structural Approximate Match:} For elements lacking unique semantic markers, we retain the stable XPath suffix, preserving local structural context while tolerating orientation-induced changes near the root hierarchy.
    \end{enumerate}

If the target widget is not immediately visible (e.g., due to screen bounds in the mutated landscape orientation), the execution engine triggers an exploratory scrolling mechanism. It iteratively performs scroll actions and re-evaluates the hierarchy until the target widget is located or a scroll limit is reached. This adaptive approach ensures high execution success rates even when layout adaptability alters component positions.

\subsection{VLM-based Bug Detection}

As the downstream module of RotDroid, this component relies on the RotVL model to detect and localize defects.
The process initiates with the Bug Detection phase, where the portrait--landscape pair is encapsulated into a detection prompt defined in Task I in Figure~\ref{fig:prompt}. 
The model outputs a boolean verdict indicating the presence of a bug. 
Upon identifying a defect, the system triggers the secondary Bug Classification and Localization phase. 
By utilizing the classification and localization prompt defined in Task II of Figure~\ref{fig:prompt}, the model infers structured defect attributes, including the defect type, the affected screen orientation, and its 2D spatial location.
Finally, the system generates a comprehensive bug report. This includes: (1) the defective screenshots with annotated bounding boxes and class labels based on the model's prediction, and (2) the full execution trace from the SPS execution phase, ensuring the issue is reproducible.

\section{Evaluation}
\label{sec:evaluation}
In this section, we evaluate RotDroid by answering the following research questions.  
\begin{itemize}
\item RQ1: How effective is RotVL in detecting GUI rotation bugs? 
\item RQ2: How does RotDroid compare with existing data loss detection tools?
\item RQ3: How does RotDroid compare with existing orientation change-induced GUI failures detection tools?
\item RQ4: Can RotDroid detect unknown rotation bugs in real-world open-source applications? 
\item RQ5: Can RotDroid detect unknown rotation bugs in real-world closed-source applications? 
\end{itemize}
RQ1 investigates the core capability of RotVL on cross-orientation bug detection, classification, and localization. RQ2 evaluates whether RotDroid can detect rotation-induced state restoration failures more effectively than representative data loss detection tools. RQ3 compares RotDroid with prior DOC same-orientation checking, which validates consistency by comparing the initial and final portrait states after two consecutive rotations. RQ4 and RQ5 evaluate the practical usefulness of RotDroid on open-source and closed-source Android apps, respectively.

\subsection{Implementation}
\label{subsubsec:implementation}
We leverage \texttt{Apktool} \cite{apktool} for resource extraction, followed by \texttt{zipalign} \cite{zipalign} and \texttt{apksigner} \cite{apksigner} to repackage the instrumented applications. For dynamic data acquisition, we utilize the \texttt{uiautomator2} \cite{uiautomator2} framework to capture synchronized screenshots and view hierarchies.

The model training process was conducted on NVIDIA V100 GPUs. We leveraged the LLaMA-Factory~\cite{zheng2024llamafactory} framework to execute the fine-tuning pipeline. To comprehensively evaluate the scalability of our approach, we developed the RotVL series across three different parameter scales: RotVL-2B, RotVL-4B, and RotVL-8B. These models were fine-tuned from the official Qwen3-VL-2B-Instruct, Qwen3-VL-4B-Instruct, and Qwen3-VL-8B-Instruct checkpoints, respectively. 
The maximum training duration was set to 5 epochs, with an early stopping mechanism triggered if the validation loss did not improve for 3 consecutive evaluation steps.

RotDroid runs offline on a host PC with 32 GB memory and a 1 TB SSD, while RotVL inference is served through an API on one NVIDIA V100 GPU. The test device only provides screenshots and view hierarchies and executes UI events through \texttt{uiautomator2}; thus, a standard device such as Pixel 6a is sufficient. 

\begin{table*}[!ht]
\centering
\caption{Comparison of RotVL with Baselines on RotBench Dataset.}
\label{tab:rotvl}
\small
\setlength{\tabcolsep}{3pt} 
\renewcommand{\arraystretch}{1.1} 
\begin{tabular*}{\textwidth}{@{\extracolsep{\fill}}lcccccccccccccc}
\toprule
\multirow{2}{*}{Model}
& \multicolumn{4}{c}{Bug Detection   (\%)}
& \multicolumn{4}{c}{Bug Classification (\%)}
& \multicolumn{4}{c}{Bug Loc. (Orient.) (\%)}
& \multicolumn{2}{c}{Bug Loc. (Coord.)} \\
\cmidrule(lr){2-5} \cmidrule(lr){6-9} \cmidrule(lr){10-13} \cmidrule(lr){14-15}
& Acc. & Prec. & Rec. & F1 & Acc. & Prec. & Rec. & F1 & Acc. & Prec. & Rec. & F1 & Dist. (px) & Ratio \\
\midrule
Qwen3-VL-2B
& 53.24 & 62.82 & 53.24 & 42.48 & 20.59 & 8.76 & 20.59 & 9.56 & 54.12 & 54.41 & 54.12 & 53.34 & 466 & 73.29 \\

Qwen3-VL-4B
& 61.47 & 61.54 & 61.47 & 61.41 & 21.47 & 20.54 & 21.47 & 13.03 & 52.06 & 75.53 & 52.06 & 37.75 & 499 & 53.72 \\

Qwen3-VL-8B
& 66.47 & 67.65 & 66.47 & 65.90 & 30.59 & 52.49 & 30.59 & 26.02 & 54.41 & 69.03 & 54.41 & 43.58 & 435 & 38.45 \\

Qwen3-VL-32B
& 58.82 & 59.73 & 58.82 & 57.84 & 30.00 & 65.02 & 30.00 & 25.89 & 53.82 & 75.99 & 53.82 & 41.31 & 478 & 30.07 \\

Qwen3-VL-235B
& 60.00 & 60.72 & 60.00 & 59.32 & 37.06 & 46.05 & 37.06 & 33.77 & 56.18 & 74.48 & 56.18 & 46.10 & 459 & 45.06 \\

GPT-5.2
& 70.88 & 71.01 & 70.88 & 70.84 & 49.71 & 65.70 & 49.71 & 49.95 & 68.82 & 70.61 & 68.82 &68.13 & 276 & 9.76 \\

\midrule
RotVL-2B
& 71.76 & 74.21 & 71.76 & 71.03 & 48.53 & 59.51 & 48.53 & 42.97 & 57.35 & 68.95 & 57.35 & 49.56 & 237 & 4.87 \\

RotVL-4B
& 76.47 & 76.50 & 76.47 & 76.46 & 44.41 & 65.37 & 44.41 & 39.54 & 71.47 & 77.38 & 71.47 & 69.84 & 263 & 16.62 \\

\textbf{RotVL-8B}
& \textbf{85.29} & \textbf{85.37} & \textbf{85.29} & \textbf{85.29} 
& \textbf{62.65} & \textbf{66.71} & \textbf{62.65} & \textbf{62.68} 
& \textbf{80.29} & \textbf{82.36} & \textbf{80.29} & \textbf{79.97} 
& \textbf{198} & \textbf{3.17} \\
\bottomrule
\end{tabular*}
\end{table*}

\subsection{Evaluation Setup}
\subsubsection{Datasets}
To comprehensively evaluate the effectiveness of RotDroid, we constructed six distinct datasets tailored to our research questions:

\paragraph{RotBench Dataset for RQ1} 
We utilize the testing part of the RotBench dataset introduced in Section~\ref{subsec:RotBench}, with detailed statistics provided in Table~\ref{tab:rq1}.
This testing dataset ensures zero leakage of project-specific data, providing a rigorous benchmark for evaluating the model's generalization capability on unseen applications.
\paragraph{Natural Bug Set for RQ1}
To further assess whether RotVL generalizes beyond synthetic mutations, we constructed a natural bug set from real-world confirmed bugs. Specifically, we randomly sampled 100 pairs of runtime cross-orientation screenshots from 44 real-world apps whose rotation bugs had been confirmed by developers, and manually annotated them for bug detection.
\paragraph{Dataset for RQ2} 
We compare RotDroid against two representative state-loss detection tools, iFixDataloss~\cite{guo2022ifixdataloss} and DLD~\cite{riganelli2020data-dld}. Since the two tools were evaluated on partially overlapping app collections, we use the intersection of their published datasets as the comparison benchmark. This results in a set of 48 Android applications. 

\paragraph{Dataset for RQ3}
Since the implementation of the DOC-based approach in~\cite{amalfitano2018does} is not fully publicly available and it generates the initial set of test cases manually, we reproduced its detection module following the description in the paper, while using RotDroid’s exploration module to drive the generation of the initial test cases. 
For the benchmark, we collected the open-source apps with the same versions as those reported in~\cite{amalfitano2018does}. 
Based on package names, version information, and linked repositories, we retrieved historical APKs from IzzyOnDroid~\cite{izzyondroid}, F-Droid, and GitHub releases. 
In total, corresponding-version APKs are successfully downloaded for 65 of the 68 apps reported in~\cite{amalfitano2018does}, while the remaining 3 APKs are unavailable.

For RQ2 and RQ3, RotDroid and all baseline methods were evaluated under the same budget, with 30 minutes of testing per app.

\paragraph{F-Droid Dataset for RQ4} This dataset consists of open-source applications collected from the F-Droid repository, as initially introduced in Section~\ref{sec:RotVL}. We perform filtering on the initial collection by excluding apps that could not be launched via ADB or did not support screen rotation. Finally, we randomly selected 300 open-source applications from the resulting pool.

\paragraph{Google Play Dataset for RQ5} To evaluate RotDroid on closed-source, commercial software, we crawled the top 20 apps from each Android application category on Google Play. After filtering out a random subset and removing applications that were incompatible with ADB or lacked rotation support, we arrived at a final set of 103 closed-source applications.

\subsubsection{Evaluation Metrics}
\paragraph{Metrics for RQ1}
To comprehensively assess the effectiveness of the RotVL model, we evaluate performance across three dimensions: \textit{Bug Detection}, \textit{Bug Classification}, and \textit{Bug Localization}. 
Given that localization requires identifying the target image before pinpointing the region, we divide the localization task into two sub-tasks: \textit{Bug Localization (Orientation)} (i.e., identifying whether the bug is in portrait or landscape mode) and \textit{Bug Localization (Coordinate)} (i.e., identifying the spatial bounding box). For the RotBench test set, we evaluate all three tasks, while for the natural bug set we report only bug detection metrics, as it is annotated only for bug detection.

We employ Accuracy, Precision, Recall, and F1-score to evaluate the model's performance on \textit{Bug Detection}, \textit{Bug Classification}, and \textit{Bug Localization (Orientation)}.

We utilize two spatial measures to evaluate \textit{Bug Localization (Coordinate)}.
\begin{itemize}
    \item \textbf{Center Point Distance:} The Euclidean distance in pixels between the predicted center $(x_{pred}, y_{pred})$ and the ground truth center $(x_{gt}, y_{gt})$, calculated as $D = \sqrt{\Delta x^2 + \Delta y^2}$, where $\Delta x = x_{pred} - x_{gt}$ and $\Delta y = y_{pred} - y_{gt}$.
    \item \textbf{Area Ratio:} The ratio of the predicted bounding box area to the ground truth area, used to measure the size accuracy of the prediction, calculated as $\frac{Area_{pred}}{Area_{gt}}$.
    \end{itemize}
    
\paragraph{Metrics for RQ2 \& RQ3}
We report \textbf{T}, \textbf{TP}, and \textbf{FP}, denoting the total number of detected bugs, true positives, and false positives, respectively.

\paragraph{Metrics for RQ4 \& RQ5}
We report (1) \textbf{Submitted Issues}, the number of issues submitted to developers based on RotDroid's findings, and (2) \textbf{Verification Status}, including whether each issue was confirmed, fixed, pending, or rejected.

\subsection{RQ1: Effectiveness of RotVL}
\label{sec:rq1}
To answer RQ1, we systematically evaluate RotVL on the testing part of RotBench dataset across three core tasks: Bug Detection, Bug Classification, and Bug Localization. Table~\ref{tab:rotvl} presents the comprehensive results. Our analysis focuses on three key dimensions: (1) improvements over base models across scales (2B, 4B, 8B), (2) comparison with massive open-source models (Qwen3-VL-32B/235B), and (3) performance against the closed-source baseline GPT-5.2.

For bug detection, RotVL consistently improves over its corresponding Qwen3-VL base models across all scales. 
RotVL-8B achieves an F1-score of 85.29\%, establishing a new state-of-the-art. 
It surpasses the large open-source baselines Qwen3-VL-32B (57.84\%) and Qwen3-VL-235B (59.32\%) by over 25\%, and also outperforms GPT-5.2 (70.84\%) by 14.45\%. 
It is worth noting that even our smaller models, RotVL-2B and RotVL-4B, surpass all open-source and closed-source baselines in Accuracy, Precision, Recall, and F1-score, proving that smaller, specialized models can effectively outperform larger general models for the GUI rotation bug detection task.

For bug classification, RotVL-8B shows stronger classification capability, achieving an F1-score of 62.68\%. It exceeds Qwen3-VL-32B (25.89\%), Qwen3-VL-235B (33.77\%), and GPT-5.2 (49.95\%) by 36.79\%, 28.91\%, and 12.73\%, respectively. Fine-tuning is again highly effective for smaller models. For example, RotVL-2B improves from 9.56\% to 42.97\% in F1, even surpassing Qwen3-VL-235B. 

For bug localization, RotVL-8B achieves the best results on both orientation and coordinate prediction. In Bug Localization (Orientation), it reaches an F1-score of 79.97\%, outperforming GPT-5.2 (68.13\%) by 11.84 points and Qwen3-VL-235B (46.10\%) by 33.87\%; RotVL-4B already exceeds GPT-5.2, and RotVL-2B also performs better than Qwen3-VL-235B. In Bug Localization (Coordinate), RotVL-8B obtains the lowest Center Point Distance of 198px, compared with 459px for Qwen3-VL-235B and 276px for GPT-5.2, and achieves the best Area Ratio of 3.17, much closer to the ideal value of 1. Similar improvements are also observed for smaller variants, with both RotVL-2B and RotVL-4B substantially reducing localization error relative to their base models.

We further conducted significance testing between RotVL-8B and the strongest baseline. Using McNemar's test~\cite{mcnemar1947note} for discrete metrics and the Wilcoxon signed-rank test~\cite{wilcoxon1992individual} for continuous metrics, the improvements of RotVL-8B are statistically significant across all evaluated dimensions: Bug Detection, Bug Classification, Orientation Localization, Center Point Distance, and Area Ratio Error, all with $p < 0.001$.

\begin{table}[!ht]
\centering
\caption{Bug detection results on the natural bug set.}
\label{tab:naturalbug}
\small
\renewcommand{\arraystretch}{1.1}
\begin{tabular}{lcccc}
\toprule
Model & Acc. & Prec. & Rec. & F1 \\
\midrule
Qwen3-VL-8B & 63.00 & 70.22 & 63.00 & 66.41 \\
GPT-5.2 & 65.00 & 59.21 & 65.00 & 61.97 \\
\textbf{RotVL-8B} & \textbf{70.00} & \textbf{74.81} & \textbf{70.00} & \textbf{72.32} \\
\bottomrule
\end{tabular}
\end{table}

To further evaluate generalization on real-world bugs, we additionally tested the models on the natural bug set. As shown in Table~\ref{tab:naturalbug}, RotVL-8B also achieves the best performance on this dataset, with an Accuracy of 70.00\%, Precision of 74.81\%, Recall of 70.00\%, and F1-score of 72.32\%, outperforming both Qwen3-VL-8B and GPT-5.2. These results suggest that RotVL does not merely learn synthetic mutation patterns from RotBench, but also generalizes to naturally occurring GUI rotation bugs in real applications.

Overall, RotVL-8B achieves the best results across bug detection, classification, and localization. 
Comparisons with the corresponding Qwen3-VL backbones serve as a fine-tuning ablation and show that rotation-specific tuning better distinguishes valid responsive adaptations from user-visible rotation-induced inconsistencies.

\begin{tcolorbox}[
    colback=gray!10,
    colframe=black,
    arc=4pt,
    boxrule=1pt,
    fonttitle=\bfseries
]
\textbf{Answer to RQ1:}
RotVL-8B achieves the best performance across all GUI rotation bug diagnosis tasks, reaching 85.29\% F1 in bug detection and outperforming both large open-source VLMs and GPT-5.2.
\end{tcolorbox}

\subsection{RQ2: Comparison with Data Loss Detection Tools}
\begin{table}[!ht]
\centering
\caption{Comparison with data loss detection tools.}
\label{tab:rq2}
\small
\setlength{\tabcolsep}{4pt}
\renewcommand{\arraystretch}{1}
\begin{tabular}{lccccccc}
\toprule
\textbf{Methods} & \multicolumn{3}{c}{RotDroid} & \multicolumn{3}{c}{DLD} & iFixDataloss \\
\cmidrule(lr){2-4} \cmidrule(lr){5-7} \cmidrule(lr){8-8}
\textbf{Metrics} & T & TP & FP & T & TP & FP& T \\
\midrule
\textbf{\#Unique Bugs} & 354 & \textbf{300} & 54 & 219& 167 & 52 & 125 \\
\bottomrule
\end{tabular}
\end{table}
As shown in Table~\ref{tab:rq2}, RotDroid detected 300 true bugs, substantially outperforming DLD (167) and iFixDataloss. RotDroid also achieves a precision of 84.75\%, compared with 76.26\% for DLD.
These results indicate that RotDroid is more effective at exposing rotation-induced GUI failures than existing data loss detection tools.

We attribute this improvement to two design differences. First, iFixDataloss and DLD mainly focus on state variable loss, whereas RotDroid reasons about cross-orientation GUI equivalence and can therefore capture failures that manifest as user-visible state inconsistency after rotation. Second, RotDroid systematically constructs mutated state-preserving sequences, which helps expose state restoration failures that occur only when rotation is interleaved with normal GUI interactions.

Most false positives are caused by dynamic or transient GUI content. For example, apps displaying clocks or dynamically changing game content may naturally show different states across orientations. Some false positives are also introduced by temporary UI elements, such as blinking input cursors or Autofill/Paste floating contextual menus that disappear after a few seconds. In addition, some landscape views display extra icons or layout elements as part of normal responsive adaptation, which may be mistakenly reported as inconsistencies.

\begin{tcolorbox}[
    colback=gray!10, 
    colframe=black, 
    arc=4pt,
    boxrule=1pt, 
    fonttitle=\bfseries 
]
\textbf{Answer to RQ2:}
RotDroid detects more true rotation-induced state restoration bugs than representative data-loss detection tools, achieving 300 true positives with 84.75\% precision. This shows the benefit of cross-orientation GUI reasoning for detecting state restoration failures.
\end{tcolorbox}

\subsection{RQ3: Comparison with a Rotation-based GUI Failure Detection Method}
\begin{table}[!ht]
\centering
\caption{Comparison with the DOC-based method.}
\label{tab:rq3}
\small
\renewcommand{\arraystretch}{1}
\begin{tabular}{lcccccc}
\toprule
\textbf{Methods} & \multicolumn{3}{c}{RotDroid} & \multicolumn{3}{c}{DOC} \\
\cmidrule(lr){2-4} \cmidrule(lr){5-7}
\textbf{Metrics} & T & TP & FP & T& TP & FP \\
\midrule
\textbf{\#Unique Bugs} & 318 & \textbf{268} & 50 & 161 &139 &22 \\
\bottomrule
\end{tabular}
\end{table}

Table~\ref{tab:rq3} presents the results. RotDroid detected 268 true bugs, substantially outperforming the DOC baseline (139). RotDroid achieved a precision of 84.28\%, compared with 86.34\% for DOC. Although DOC has slightly higher precision, RotDroid detects nearly twice as many true bugs, demonstrating stronger overall detection capability.

The DOC baseline can be viewed as an ablation without SPS mutation and cross-orientation consistency checking: it validates same-orientation consistency by comparing the initial and final portrait states after a portrait$\rightarrow$landscape$\rightarrow$portrait sequence. This design can detect failures that persist after the second rotation, but may miss failures that appear only in the intermediate landscape state and disappear after rotating back. RotDroid avoids this limitation by directly comparing portrait--landscape state-equivalent pairs through RotVL.

\begin{tcolorbox}[
    colback=gray!10,
    colframe=black,
    arc=4pt,
    boxrule=1pt,
    fonttitle=\bfseries
]
\textbf{Answer to RQ3:}
RotDroid detects more true GUI rotation bugs than the DOC-based method, achieving 268 true positives with 84.28\% precision. This shows the advantage of directly checking portrait--landscape state equivalence.
\end{tcolorbox}

\subsection{RQ4: Effectiveness on Real-world Open-source Apps}

We evaluated RotDroid on a diverse set of open-source Android apps. In total, RotDroid identified 78 GUI rotation issues. Among them, 72 were submitted to developers via GitHub Issues, while the remaining 6 were reproducible in the tested versions but had already been fixed in newer releases. 

\begin{table}[!ht]
\centering
\caption{Distribution of identified GUI issues in open-source Android applications by bug category.}
\label{tab:bugs-open}
\small
\setlength{\tabcolsep}{5pt}
\renewcommand{\arraystretch}{1.1}
\begin{tabular}{lccc}
\toprule
\textbf{Bug Category} & \textbf{Confirmed} & \textbf{Pending} & \textbf{Rejected} \\
\midrule
Layout Failures & 21 & 13 &0  \\
State Restoration Failures & 25 & 15 & 1 \\
Rotation Execution Failures & 1 & 3 & 0 \\
\midrule
\textbf{Total} & 47 & 30 & 1 \\
\bottomrule
\end{tabular}
\end{table}

As shown in Table~\ref{tab:bugs-open}, 47 issues have been confirmed, including 29 fixed issues, 12 confirmed but not yet fixed issues, and 6 issues that were already fixed in newer versions; 30 issues are still pending, and 1 was rejected.
Among the confirmed issues, 21 are layout failures, 25 are state restoration failures, and 1 is a rotation execution failure. Rotation execution failures are relatively rare because our evaluation excludes apps that do not support rotation and focuses on apps where portrait--landscape state-equivalent pairs can be constructed. 
The rejected report was reproducible in our environment but could not be reproduced by the developer, suggesting that some results may depend on device or runtime environments. Overall, these results show that RotDroid can effectively detect diverse real-world rotation issues in open-source Android apps.

\paragraph{Developer Feedback}
The utility of RotDroid was further evidenced by developer feedback. Developers actively engaged with our reports, often providing fixes or constructive discussions. 
For example, a developer of \textit{Infomaniak kMeet} responded: \textit{``Your issue has been Fixed and merged into the main branch and will be released very soon with the next release. Scrolling is now possible to access the missing buttons. Thank you again!''}
Regarding the issue found in \textit{DuckDuckGo}, a contributor commented on the complexity of handling orientation changes: \textit{``I think that's a great question for an interview for an Android Engineer, on how to handle screen orientation changes \texttwemoji{rolling on the floor laughing}''}
These interactions indicate that RotDroid reports are useful to developers and help improve rotation robustness.

\paragraph{Industrial Relevance}
The detected issues also have practical impact. RotDroid reported bugs in widely used open-source apps, including \textit{DuckDuckGo}, which has 50M+ installs. Some apps with relatively few repository stars, such as \textit{Infomaniak kMeet} and \textit{Simple Crypto Widget}, still have 100k+ installs on Google Play. This shows that RotDroid can uncover rotation bugs in real applications with substantial user bases, beyond small or toy projects.

\begin{tcolorbox}[
    colback=gray!10,
    colframe=black,    
    arc=4pt, 
    boxrule=1pt,
    fonttitle=\bfseries 
]
\textbf{Answer to RQ4:} 
RotDroid is effective in real-world open-source applications, with 41 confirmed issues among 72 submitted GUI rotation issues and positive developer feedback.
\end{tcolorbox}

\subsection{RQ5: Effectiveness on Real-world Closed-source Apps}
\begin{table}[!ht]
\centering
\caption{Closed-source Android applications with identified GUI issues.}
\label{tab:bugs-close}
\scriptsize
\setlength{\tabcolsep}{3pt}
\renewcommand{\arraystretch}{1}
\begin{tabular}{lc|lc|lc|lc|lc}
\toprule
\#1  & Dropbox   & \#5  & ibis*     & \#9  & Tasks    & \#13 & MANGA*   & \#17 & Toomics \\
\#2  & Phone*    & \#6  & ReadEra   & \#10 & Canvas   & \#14 & Yr       & \#18 & Zillow  \\
\#3  & Health    & \#7  & Infinite* & \#11 & Walli    & \#15 & Cabify*  & \#19 & US News  \\
\#4  & Lark*     & \#8  & Cars*     & \#12 & Trulia   & \#16 & Fuelio   & \#20 & OurFa*   \\
\bottomrule
\multicolumn{10}{l}{\scriptsize * denotes abbreviated application names.}
\end{tabular}
\end{table}

To assess the generalizability of RotDroid beyond open-source repositories, we evaluated it on closed-source Android applications from Google Play. Table~\ref{tab:bugs-close} lists 20 representative applications in which RotDroid identified 22 GUI rotation issues, 6 of which have been confirmed.
These applications span 15 diverse categories, suggesting that RotDroid is not limited to specific UI patterns or app types. Notably, RotDroid detected issues in widely used commercial apps such as \textit{Dropbox} (\#1), \textit{Phone by Google} (\#2), and \textit{Samsung Health} (\#3), each with over one billion installs. It also found issues in popular apps such as \textit{Lark Player} (\#4) and \textit{ibis Paint X} (\#5), which have over 100 million installs. These results indicate that GUI rotation bugs can remain overlooked even in mature commercial applications.

\begin{tcolorbox}[
    colback=gray!10, 
    colframe=black, 
    arc=4pt, 
    boxrule=1pt, 
    fonttitle=\bfseries 
]
\textbf{Answer to RQ5: }RotDroid is effective on closed-source Android apps, detecting 22 GUI rotation issues in 20 representative commercial applications, including several with hundreds of millions or even billions of installs.
\end{tcolorbox}

\section{Case Studies and Discussion}
\label{sec:case study}

\subsection{Layout Failures in Florae}
\begin{figure}[htbp]
\centering
\includegraphics[width=0.9\columnwidth]{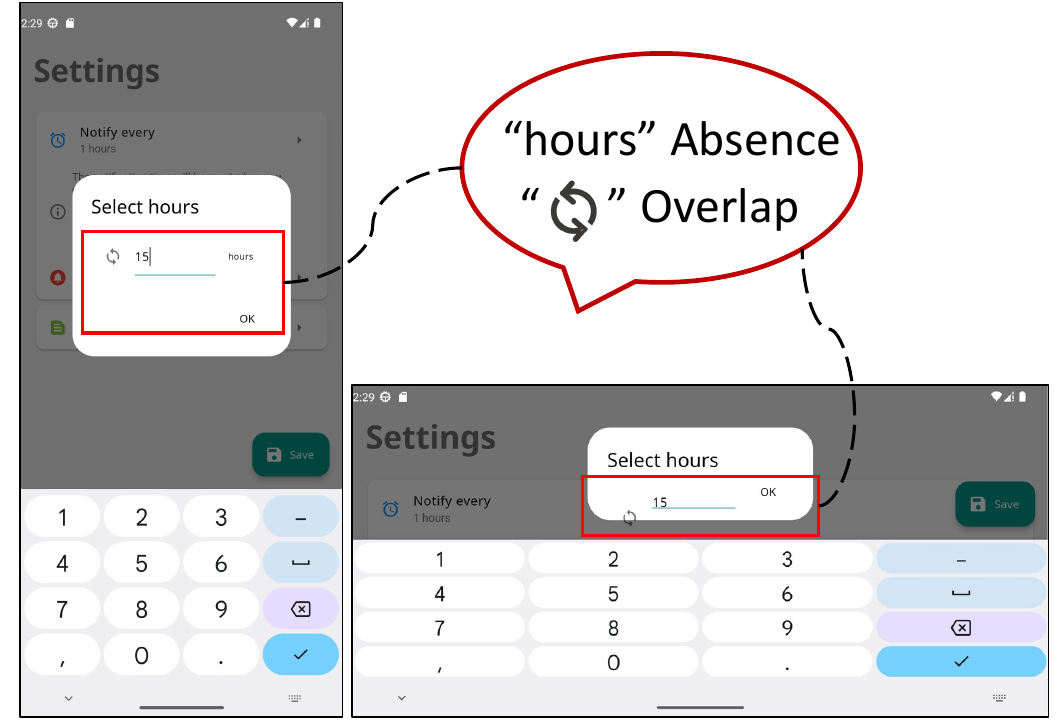}
\caption{GUI Layout Failure in Florae.}
\label{fig:case}
\end{figure}
As shown in Figure \ref{fig:case}, we identify a layout failure in Florae from two cross-orientation views that are expected to be state-equivalent. In the portrait view, the ``Select hours'' dialog is rendered correctly: the input value, the unit label ``hours'', the reset icon, and the ``OK'' button are all visible and spatially separated. 
However, in the corresponding landscape view, the dialog fails to properly reposition its internal components under the new orientation.
The ``hours'' label is completely occluded by the ``OK'' button, resulting in a component absence effect, while the reset icon overlaps with the boundary of the ``Select hours'' area. RotDroid detects this bug by executing a mutated state-preserving action sequence and comparing the portrait and landscape states that should remain semantically equivalent. This case cannot be detected by approaches that only check whether the portrait state before and after two consecutive rotations remains consistent, because the portrait layout itself is restored correctly and the failure exists only in the intermediate landscape state.

\subsection{DuckDuckGo Welcome Flow Reset on Rotation}

\begin{figure}[htbp]
\centering
\includegraphics[width=1\columnwidth]{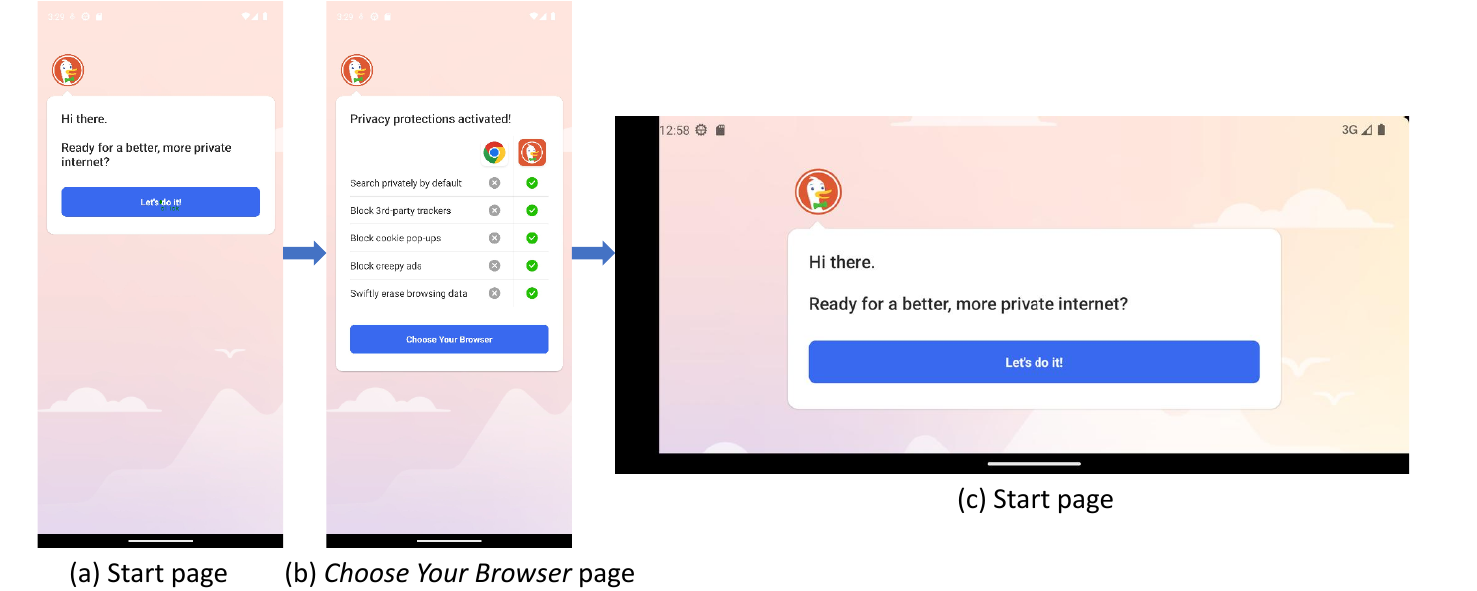}
\caption{GUI State Restoration Failure in DuckDuckGo.}
\label{fig:duckduckgo}
\end{figure}
As shown in Figure \ref{fig:duckduckgo}, RotDroid detected a state restoration bug in DuckDuckGo's onboarding flow. After the user tapped \textit{``Let's do it!''} and entered the \textit{``Choose Your Browser''} page, rotating the device to landscape unexpectedly reset the flow to the start page. This bug was caused by insufficient preservation of the current onboarding step during the Android configuration change. Developers addressed the issue with two alternative fixes: one refactored the \texttt{WelcomePageViewModel} to maintain the current step in a lifecycle-aware \texttt{StateFlow}, while the other explicitly saved and restored the dialog type through \texttt{onSaveInstanceState}. This case demonstrates that RotDroid can uncover lifecycle-related rotation bugs where the layout remains valid but the application state is incorrectly restored.

\section{Threats to Validity}
Regarding \textbf{internal validity}, the synthesis of defective data may introduce inaccuracies; we therefore manually audit generated samples. The partially unavailable DOC implementation may also introduce reproduction bias, although we follow its published description and reuse corresponding-version APKs whenever available. SPS operationally preserves observable GUI state and interaction context, but cannot guarantee hidden server-side state, caches, or external side effects. Concerning \textbf{external validity}, evaluation on over 400 diverse applications supports the practical utility of RotDroid. We may further evaluate RotDroid on more commercial Android apps to strengthen our evaluation.

\section{Related Work}
\label{sec:related work}

\paragraph{Automated Mobile GUI Testing}
The landscape of Android testing has evolved from script-based frameworks \cite{zadgaonkar2013robotium,shah2014software-appium} to intelligent exploration tools. Early stochastic and model-based approaches \cite{machiry2013dynodroid,li2017droidbot,sustoat,mao2016sapienz} focused primarily on maximizing code coverage and crash detection. Recent research incorporates reinforcement learning to optimize navigation policies \cite{romdhana2022deep,guo2025effectively,yu2024effective}. To address Non-Crash Functional  defects, visual-centric methods \cite{yu2019lirat,liu2022nighthawk,liu2022woodpecker} employ computer vision to identify static display anomalies. However, these approaches typically analyze isolated screenshots and lack the mechanisms to verify dynamic layout consistency during state transitions.

\paragraph{LLM-driven Mobile GUI Understanding and Automation}
Large Language Models have significantly advanced GUI testing by enabling semantic input generation and task automation \cite{liu2023chatting-gptdroid,wen2024autodroid,liu2023fill}. Furthermore, Multimodal Large Language Models and Vision-Language Models \cite{you2024ferret,bai2025qwen3vltechnicalreport} facilitate grounded UI understanding. 
Contemporary works utilize these models for consistency checking and bug detection \cite{liu2025seeing-visiondroid,su2025automated,chen2025standing}. Recent empirical studies indicate that general-purpose VLMs still struggle with complex mobile screen understanding \cite{sravanthi2025perception} and may fail to distinguish intended responsive adaptations from diverse functional bugs without domain-specific guidance \cite{ju2024study}.

\paragraph{Configuration and Layout Adaptation Testing}
Screen rotation frequently induces both state management faults and GUI failures \cite{amalfitano2018does}. Existing research in this domain predominantly targets data persistence issues \cite{riganelli2020data-dld,guo2022ifixdataloss}. Regarding layout correctness, prior works mainly address compatibility issues arising from device fragmentation \cite{fazzini2017automated-diffdroid,ren2022cross} or screen scaling \cite{su2022metamorphosis}. However, these approaches focus on comparing static UI states in portrait mode across distinct environments and fail to capture the rendering anomalies specifically induced by the layout reflow process during orientation changes.

\section{Conclusion and Future Work}
\label{sec:conclusion}
In this paper, we present RotDroid, which combines UI Transition Graph modeling, SPS mutation, and RotVL-based cross-orientation checking to detect GUI rotation bugs. RotBench supports the domain-specific fine-tuning of RotVL, and RotDroid detects 78 bugs in open-source apps and 22 in commercial apps. Future work will extend RotDroid to video-centric applications with dynamic and orientation-sensitive content.

\section{Data Availability}
Replication package with code, model and issues: \url{https://doi.org/10.5281/zenodo.21897206}.

\section*{Acknowledgment}
This work is supported by State Key Laboratory of Complex \& Critical Software Environment (SKLCCSE).

\bibliographystyle{IEEEtran} 
\bibliography{refer}

\end{document}